%Paper: hep-th/9312163
%From: guest1@imsc.ernet.in (Math guest )
%Date: Mon, 20 Dec 93 18:55:40+050

\magnification=1200
%\raggedbottom
\hsize=4.93truein
%\vsize=20.9truecm
\vsize=7.85truein
\voffset=1.82truecm
\hoffset=1.95truecm
\nopagenumbers
\footline{\hskip13.7truecm\tenrm\folio}
\pageno=1
\overfullrule=0pt

\input macros.tex
\def\vone{\vskip12truept}
\def\vtwo{\vskip24truept}

\parskip=0pt
\parindent=3em
\baselineskip=12truept

\topinsert
 To be published in Foundations of Physics Letters
\vskip1.1truecm
\endinsert
\noindent
{\bf  PROTECTIVE MEASUREMENT AND

\noindent
QUANTUM
REALITY}
\vskip36truept
{\parindent=.8truein

{\bf J. Anandan}

{\it Department of Physics and Astronomy}

{\it University of South Carolina}

{\it Columbia, South Carolina 29208}
\vskip 12truept

\vone
Received  August 27, 1993; revised September 2, 1993

}
\vtwo

\def\lsim{<\kern-2.5ex\lower0.85ex\hbox{$\sim$}\ }
\def\rsim{>\kern-2.5ex\lower0.85ex\hbox{$\sim$}\ }

\noindent %ABSTRACT
It is shown that from the expectation values of obervables, which can
be measured for a single system using protective
measurements, the linear structure,
inner product, and observables in
the Hilbert space can be reconstructed. A universal method of
measuring the wave function of a single particle using its
gravitational field is given. Protective measurement is generalized to
the measurement of a degenerate state and to many particle
systems. The question of whether the wave function is real is
examined, and an argument of Einstein in favor of the ensemble
interpretation of quantum theory is refuted.

\vone\noindent

\vtwo\noindent
{\bf{1. INTRODUCTION}}
\vone

The measurement problem in quantum theory arises from the fact
that the wave function of a particle, in the usual Copenhagen
interpretation, collapses during the measurement to a localized wave.
e. g. When the extended wave function of a particle interacts with a
photosensitive plate or a detector, a localized spot appears. Moreover,
this outcome can only be predicted probabilistically, the wave
function then playing the role of a probability amplitude. And a
probabilistic prediction can be given physical meaning only by
associating it with an ensemble of identical particles.

Protective measurement [1,2] is a new type of quantum
measurement that enables the manifestation of the  wave function of
a {\it single} particle to be determined while it is prevented from
collapse because of another interaction which it undergoes during the
measurement. For example, the system may be in an eigenstate of
the Hamiltonian and the measurement is made adiabatically to
prevent transition to other eigenstates. Then its wave function does
not collapse, and the outcome of the measurement of the expectation
value of an observable can be predicted deterministically. By
sufficient number of such measurements, the wave function can be
observed for a single particle.

In this paper, I shall extend the meaning and scope of a protective
measurement, and discuss its implications to the reality of states and
observables in quantum theory. The protective measurement is
reviewed in Sec.~2 by means of the particular example of the
measurement of the spin state of a spin-half particle. Both ideal and
realistic protective Stern-Gerlach experiments are described. It is
shown that protectively measuring a state in an $N+1$
dimensional Hilbert space of a quantum
system amounts to determining the orientation of the vector
representing the corresponding density matrix in an $N^2+2N$
dimensional real Euclidean space.

In Sec.~3, it is shown that given the observed expectation values
as functions on the set of physical states as an abstract set, even the
existence of the Hilbert space, its inner product and the observables
as Hermitian operators can be reconstructed. Here, unlike in  the
previous treatments, the interaction which protects the state when
the expectation value is being measured is not a "secret". On the
contrary, it may be part of the definition of the state. The states
acquire an ontological meaning from the protective measurements
which determine in a non statistical manner the expectation values
of the observables for the states and therefore the inner product
between the state vectors.

A universal method for determining, in principle, the spatial part of
the wave function using the gravitational field is given in Sec.~4.
The particular example of a tube in a rotating frame considered here
leads to the extension of protective measurement for degenerate
states in Sec.~5.

In Sec.~6, it is shown how to make a protective measurement on
the wave function of a system of two particles. This is important
because the case of single particle considered before [1,2] does not
distinguish between a wave in physical 3-space and a wave in
configuration space. The extension of protective measurement to
many particle systems here clearly shows that it is the wave in
configuration space, and not physical space which Schr\"odinger
originally had in mind for the domain of the wave function, which
acquires an ontological meaning.

The question of whether the wave function is real, i. e. ontological, is
investigated in
Sec.~7. An argument due to Einstein in favor of the ensemble
interpretation is examined. This is different from the usual
arguments in favor of the statistical interpretation in that it does not
involve the collapse of the wave function. But it is concluded that the
possibility of protectively measuring the wave function, unknown to
Einstein, invalidates this argument.

\vtwo\noindent
{\bf{2. PHYSICS AND GEOMETRY OF A PROTECTIVE

\parindent=1.25em

MEASUREMENT}}
\vone

In a protective measurement the wave function of a quantum system
does not change appreciably while the measurement is being made
on it.  One way of doing this is by having the system undergo a
suitable interaction so that it is in a non degenerate eigenstate of the
Hamiltonian. Then the measurement is made adiabatically so that the
wave function  neither changes appreciably nor become entangled
with the apparatus [1, 2]. Therefore the wave function does not
collapse during the measurement as in the usual measurement. The
suitable interaction that the system undergoes which enables such a
measurement to be made will be called the protection, because it
prevents the wavefunction from undergoing a transition from the
given energy eigenstate into another energy eigenstate. And thereby
it protects the wave function from entanglement or collapse.

Suppose the Hamiltonian  of the combined system consisting of the
given system and the apparatus is
$$H = H_0+g(t)qA+H_a \eqno(2.1)$$
where $H_0$ is the Hamiltonian of the given  system when it is not
interacting with the apparatus, but including the protection. $H_a$ is
the Hamiltonian of the apparatus, $A$ is the observable of the
system being measured which is coupled to an observable $q$ of the
apparatus, and g(t) is a c-number function of time which represents
the turning on and off of the interaction between the system and
apparatus during the interval $[0,T]$. Indeed, the interaction
Hamiltonian $g(t)qA$ may be used to provide a prescription for
associating an observable with the apparatus: Once the scale of
$g(t)q$ is fixed by convention, the observable $A$ to be associated
with this apparatus interacting with the system is uniquely
determined by the interaction Hamiltonian.

The time of measurement $T$ is roughly the time taken for the
change in the typical matrix elements of the Hamiltonian $H$ to be of
the order of the smallest energy difference $\delta E$ between
the given state and other eigenstates of energy.  By this I mean,
$$T \sim {\delta E\over {\langle{{\partial H}\over {\partial
t}}\rangle}}.\eqno(2.2)$$
Then  the condition for adiabaticity is
$T>>{\hbar \over \delta E} $. i. e.
$$\hbar {{\langle{{\partial H}\over {\partial t}}\rangle}\over \delta
E^2}<<1 \eqno(2.3)$$
This is the same as the condition for the transition from the given
state to other eigenstates of energy to be negligible [3]. Therefore the
system continues to remain as an eigenstate of the Hamiltonian:
$$|\psi (t)> = exp\{-{i\over \hbar} \int\limits_0^t E(t')dt'\} |E(t)>,
\eqno(2.4)$$
where $|E(t)>$ is the eigenstate of the Hamiltonian belonging to the
eigenvalue $E(t)$. Hence this state does not become
entangled with the apparatus state. Consequently the state does not
collapse during the measurement which avoids the usual
interpretation of the wave function in terms of the probabilities of
the possible outcomes of the collapse.

Assume also that the interaction Hamiltonian is small compared to
the unperturbed Hamiltonian. Then $|\psi>$ does not change
appreciably because of the interaction so that the state being
measured is approximately the same as the original state. Also, first
order perturbation theory then gives, for any value of q [1,3]
$$E(t) = E_o(t) + g(t) q<\psi(t)|A|\psi(t)> , \eqno(2.5)$$
where $E_o(t)$ is the eigenvalue of the Hamiltonian in the absence of
the interaction. It follows from (2.4) and (2.5) that, as a result of the
interaction, there is an exponential factor acting on the wave function
of the apparatus which changes the variable $p$ conjugate to $q$ by
$$\Delta p = -<\psi|A|\psi> . \eqno(2.6)$$
where for inessential simplicity the normalization
$$\int\limits_0^T g(t)dt = 1 \eqno(2.7)$$
is assumed.
Therefore by measuring $\Delta p$ for the apparatus, the expectation
value of the observable $<\psi|A|\psi>$ is
obtained for a single particle. By measuring the expectation values in
this manner for sufficient number of observables, $|\psi>$ can be
reconstructed up to an overall phase and normalization.

This is most easily illustrated by measuring the spin state of a
neutral spin $1\over 2$
particle such as a neutron or a suitable atom with a magnetic
moment. Suppose $\bf m$ is a unit vector such that
$${\bf \sigma}\cdot {\bf m}| \psi \rangle = | \psi \rangle ,
\eqno(2.8)$$
where $\sigma ^i,i=1,2,3$ are the Pauli spin matrices. Then we shall
say that the spin of the state $| \psi \rangle$ is in the
direction of ${\bf m}$. Suppose that we do not know the spin state of
a particle, but know only that it is protected. i. e. there is a large
magnetic field $B_o$ in the unknown direction of the spin. In the
next section, even this requirement that the protecting magnetic field
is unknown will be removed. The objective now is to determine the
spin state by measurements on this particle alone. This is unlike the
usual measurement of a quantum mechanical state by measurements
on an ensemble of identical systems in that state. Determining the
spin state up to phase and normalization is the same as determining
the unit vector $\bf m$ defined above.

Send the particle, which has magnetic moment $\mu$ through an
ideal Stern-Gerlach apparatus whose magnetic field and its
inhomogeneity are in the direction of the unit vector ${\bf n}_1$. i. e.
in (2.1), $A=\mu {\bf \sigma}\cdot {\bf n}_1$ with q being the
coordinate of the center of mass of the particle, which plays the role
of the apparatus, in the direction of ${\bf n}_1$. Then, because of the
protection, the wave function does not split into two as in the usual
Stern-Gerlach experiment [2]. But its center of mass undergoes an
acceleration which can be observed by shining a laser beam on the
particle. It follows from (2.6) that the change in momentum due to
this acceleration is proportional to
$$\langle \psi|{\bf \sigma}\cdot {\bf n}_1| \psi \rangle =
{\bf m}\cdot {\bf n}_1\eqno(2.9)$$
 provided $|\psi \rangle$ is normalized.

Therefore from this acceleration the angle between $\bf m$ and
${\bf n}_1$ can be determined. Now send this particle through
another ideal Stern-Gerlach apparatus in which the inhomogeneity of
the magnetic field is in the direction of ${\bf n}_2$. From the
acceleration we can determine ${\bf m}\cdot {\bf n}_2$. Now there
are only two possible values for ${\bf m}$. To remove this discrete
ambiguity, we need to send the particle through another ideal Stern-
Gerlach apparatus in which the inhomogeneity of the magnetic field
is along ${\bf n}_3$ such that ${\bf n}_1$, ${\bf n}_2$ and ${\bf n}_
3$ are linearly independent. From the acceleration we determine
${\bf m}\cdot {\bf n}_3$. We now know the angles that the unit
vector $\bf m$ makes with three linearly independent directions,
which uniquely determines $\bf m$. Hence, the spin state has been
determined by measurements on a single particle.

The above ideal Stern-Gerlach experiment was so named because the
magnetic field ${\bf B}$  which is proportional to $q\bf n$ does not
satisfy $div {\bf B}=0$ and is therefore fictitious. Consider now a real
magnetic field that satisfies Maxwell's equations. On doing a Taylor
expansion to first order and disregarding the constant term (which
has already been included as the protecting magnetic field
${\bf B}_0$),
we obtain $B^i=L^{ij}x^j$ using the summation convention. The
interaction term in (2.1) is then replaced by [4] $g(t)L^{ij}x^j\sigma^i$.
This is like the sum of three terms with each $x^j, j=1,2,3$ playing
the role of $q$ in (2.1). I shall assume that the constant  matrix
$L^{ij}$ is trace free and symmetric. Then
$$div {\bf B}= Tr L = 0, [curl {\bf B}]^i = -\epsilon^{ijk}L^{jk}=0.
\eqno(2.10)$$
This is the simplest
form of $\bf B$ that is inhomogeneous and satisfies Maxwell's
equations.

Now the $j$ th component of the acceleration is proportional to
$$\langle \psi|L^{ij}\sigma^i| \psi \rangle =L^{ij}m^i ,\eqno(2.11)$$
which replaces (2.9). Hence, by measuring all three components of
this acceleration, if $L^{ij}$ is non singular, $m^i$ and therefore $|\psi
\rangle$ can be determined up to phase. This is like the ideal
Stern-Gerlach experiment studied
above in which also at least three measurements are needed to
determine the spin state. But the present case has the advantage that
with one Stern-Gerlach field all three measurements can be made; it
is not necessary to send the particle through three Stern-Gerlach
fields as in the previous case. The experiment could be performed in
principle by observing the position and time of arrival of the particle
at a screen relative to its postion and earlier time of arrival
at the Stern-
Gerlach apparatus. From this the three components of the
acceleration can be computed.

Even at the final stage of the protective measurement no collapse of
the wave function occurs because only one spot is predicted by
quantum theory, which should therefore be observed. Hence, the
artificial division between the quantum system and the "classical"
apparatus to expain the collapse of the wave function in the
Copenhagen interpretation of the usual measurement is not needed
in a protective measurement.

A generalization to an $N+1$ dimensional Hilbert space ${\cal H}$,
where $N$ is a non zero integer, may be obtained easily once the
protective measurement is understood geometrically as will be
 described
now. Each ray (1-dimensional subspace) of ${\cal H}$ represents a
physical state. The set of rays  of ${\cal H}$ is the $N$-dimensional
complex projective space ${\cal P}=CP(N)$, called the projective
Hilbert space. There is a natural metric in ${\cal P}$ which arises
from
the inner product in ${\cal H}$ called the Fubini-Study metric [5]. Let
${\cal M}$ be the set of traceless Hermitian operators in ${\cal H}$.
Then
${\cal M}$ is a $N^2 + 2N$ dimensional vector space and is the Lie
algebra of the group $SU(N+1)$. There is a real Euclidean metric on
${\cal M}$ defined as follows: The inner product between any pair
$A, B\epsilon {\cal M}$  with respect to this metric is $Tr AB$.

Now any point in ${\cal P}$ may be represented uniquely
by the density
operator $\rho = |\psi \rangle \langle \psi|$ with
$|\psi \rangle$ being any normalized state in the ray in ${\cal H}$
corresponding to this point. Define the
traceless Hermitian operator
$$R(\rho) =\rho -{1\over N+1}I , \eqno (2.12)$$
where $I$ is the
identity operator. Now (2.12) defines an embedding of ${\cal P}$ into
${\cal M}$. This embedding is an isometry with respect to the
metrics mentioned above [6]. Alternatively, the Fubini-Study metric
on $\cal P$ may be {\it defined} as the metric on this embedding
induced by the above trace metric in ${\cal M}$.
This embedding is invariant under the action of
$SU(N+1)$ in ${\cal H}$. There is no proper subspace of ${\cal M}$ in
which there is such an invariant embedding.

The result of the protective measurement of any observable
$A\epsilon {\cal M}$ on any $\rho \epsilon {\cal P}$ is the inner
product
$Tr AR(\rho)$. Let $A_i , i=1,2,...,$ $N^2+2N$ form
an orthonormal basis in
${\cal M}$. Then it follows that the protective measurement of the
observables $A_i$ on $\rho$ give the Cartesian
coordinates of $R(\rho)$ in
the frame defined by {$A_i$}. This uniquely determines $\rho$. Thus
the state of a system can be reconstructed by measuring $N^2+2N$
independent observables in ${\cal H}$.

For the special case of $N=1$ considered earlier, ${\cal P}$ is a sphere
with the Fubini-Study metric being the usual metric on the sphere.
${\cal M}$ is a 3-dimensional Euclidean space spanned by the
orthonormal vectors $A_i={1\over
\sqrt2}\sigma_i, i=1,2,3$. The embedding given by (2.12) is the usual
way of regarding a sphere in a Euclidean space. The determination of
a spin
${1\over 2}$ state which is a point on this sphere by protectively
measuring $A_i$ amounts to
determining the coordinates of this point in a Cartesian coordinate
system
in the embedding Euclidean space. This geometrical
treatment makes it clear why, even though a spin ${1\over 2}$
state is
uniquely determined by two real parameters that specify the
direction of ${\bf m}$, three measurements were actually needed to
determine this state.
\vtwo\noindent
{\bf{3. NEW MEANING OF THE INNER PRODUCT

\parindent=1.25em

AND  OBSERVABLES}}
\vone

So far the inner product $\langle \phi|\psi \rangle$ between any two
normalized states $|\psi \rangle$ and $|\phi \rangle$ in the Hilbert
space ${\cal H}$ has been interpreted as the ``probability amplitude".
If a measurement is made on  $|\psi \rangle$ resulting in a
transition to $|\phi \rangle$ or other states orthogonal to $|\phi
\rangle$, then the probability of transition to  $|\phi \rangle$ is
$|\langle \phi|\psi \rangle|^2$. But to give physical meaning to
probability, an ensemble of identical systems is needed. Therefore
this interpretation of the inner product, by its very nature, associates
the inner product with an ensemble of identical systems.

But I shall show now that the protective measurement gives a new
meaning to the inner product which is associated with a single
system. Also, while the usual meaning determines only the
magnitude and not the phase of $\langle \phi|\psi \rangle$, the new
meaning will be seen to determine both the magnitude and the
phase. This will be done now by showing that from the expectation
values of Hermitian operators, which may be determined by
protective experiments on a {\it single system}, the inner product in
${\cal H}$ can be reconstructed. And this inner product is then
unique up to a
constant multiple.

Suppose that ${\cal H}$ has dimension $N+1$, where $N$ is a non
negative integer or infinity. Each point of the projective Hilbert space
${\cal P} = CP(N)$, defined as the set of rays of ${\cal H}$,
represents a
physical state of the
system. It can be determined by measuring it to be the
eigensubspace of a complete set of commuting observables or a
single observable with distinct eigenvalues, which also could be the
part of the Hamiltonian that  protects the state. Protective
measurements of a Hermitian observable $A$ determine the
expectation value
$$\alpha \equiv {\langle \psi|A|\psi \rangle\over \langle \psi|\psi
\rangle} \eqno(3.1)$$
for each
state represented by $|\psi \rangle$. Clearly, $\alpha$
is the same for
all state vectors in a given ray and therefore may be regarded as a
function on
$\cal P$ which is called the expectation value of $A$. The set of
expectation values is a vector space over the field of real numbers.
\vone

{\it Theorem.} Given the expectation values of all Hermitian
operators as
functions of the set of physical states ${\cal P}$, the vector space
structure and the inner product on the
underlying Hilbert space ${\cal H}$ can be reconstructed. The vector
space structure is unique up to
isomorphism and the inner product is unique up to multiplication
by a real positive constant. After reconstructing ${\cal H}$,
each such function on ${\cal P}$ determines uniquely the
corresponding Hermitian operator in ${\cal H}$ whose expectation
value it is.
\vone

Before proving this theorem, let us recall some useful properties of
expectation values. Consider an
arbitrary Hermitian observable $A$ with
eigenvalues $a_1,a_2,...a_{N+1}$. Then, from (3.1),
$$\alpha ={\sum_{i=1}^{N+1}|\psi_i|^2a_i \over
\sum_{i=1}^{N+1}|\psi_i|^2} $$
where $\psi_i$ are the components of $|\psi \rangle$ in an
orthogonal basis in which $A$ is diagonal. Therefore $\alpha$ is like
the center of mass coordinate of masses $|\psi_i|^2$ placed along a
line with coordinates $a_i$. Hence the value of $\alpha$ varies
between the lowest and the highest eigenvalues. Now, $\alpha$ is a
constant function on $\cal P$ if and only if all the eigenvalues of A
are equal. Consider a non constant expectation value $\alpha$. In an
infinite dimensional ${\cal H}$,
restrict further to observables for which
this function is bounded, which corresponds to the eigenvalues being
all finite.

Then the points of $\cal P$ at which $\alpha$ takes its
lowest (highest) value, which is the same as the lowest (highest)
eigenvalue of $A$, correspond to the eigenvectors belonging to this
eigenvlaue. Since the lowest and the highest eigenvalues are distinct
their eigenvectors are orthogonal. It can also be shown by
differentiating (3.1) that $\alpha$ has critical values at the points in
$\cal P$ which are eigensubspaces of $A$,  i. e. the directional
derivative of $\alpha$ in every direction is zero at these points. The
values of $\alpha$ at the critical points are the corresponding
eigenvalues.

To prove this theorem, suppose now that we do not know the vector
space ${\cal H}$ whose
rays the points of ${\cal P}$ are. We are given only the
expectation
values of observables as the real valued
functions $\{\alpha\}$ of ${\cal P}$
regarded as a set only. Then the
topological, projective, and metrical structures of ${\cal P}$ may be
reconstructed as follows. The
topology in ${\cal P}$ may be determined as the smallest topology
with
respect to which the functions {$\alpha$} are continuous. This
topology is unique.

Also, by simply looking at the values of each $\alpha$, its lowest and
highest values can be determined, even without knowing the
topology of ${\cal P}$. And a pair of points of ${\cal P}$ at which
these
two values are attained are orthogonal rays in the underlying vector
space ${\cal H}$ which will be reconstructed. Two orthogonal rays in
${\cal H}$ will be called orthogonal points in ${\cal P}$. A remark
irrelevant to this proof is the following: The lowest (highest)
eigenvalue is attained exactly at one point on $\cal P$ if and only if
this eigenvalue is non degenerate. But the above treatment  is for the
more general case when these two eigenvalues may be degenerate.
For any pair of orthogonal rays there exists a Hermitian operator
whose eigenspaces belonging to the lowest and the highest
eigenvalues are these two rays. Therefore by the above procedure all
pairs of orthogonal points in $\cal P$ may be determined.

If the maximum number of mutually orthogonal points is $N+1$, then
$\cal P$ has complex dimension $N$. Consider any set $U$ of $k$
mutally orthogonal points in
$\cal P$, where $k \le N$ is a positive integer or infinity. Let $V$ be
the set of points in $\cal P$ that are
orthogonal to every
point is $U$. Then the linear space in
$\cal P$, uniquely associated with $U$, is defined
to be the set of points that are orthogonal to every point in
$V$. This space has complex dimension $k-1$. Since all pairs
of orthogonal points are known, we know also all the linear spaces in
$\cal P$. This determines uniquely the projective geometry on
$\cal P$ [7]. This projective geometry determines a vector space
$\cal H$ of
dimension $N+1$, unique up to isomorphism, such that its rays are
the points of $\cal P$. (The components in some basis of a set of
vectors in ${\cal H}$, with one chosen from each ray in a
differentiable
manner, serve as homogeneous coordinates in ${\cal P}$.)

Two vectors belong to two orthogonal points in ${\cal P}$ if and only
if they are orthogonal with respect to
the inner product in ${\cal H}$, which will now be
reconstructed. Choose in $\cal H$ a mutually
orthogonal set of vectors
$\{|\psi_1 \rangle,|\psi_2 \rangle,...|\psi_{N+1}\rangle\}$.
Now,  $|\psi_1 \rangle+b|\psi_i \rangle, i\ne1, b\ne0$,
is in the subspace spanned
by  $|\psi_1 \rangle$ and $|\psi_i \rangle$. Therefore it must have a
non null orthogonal vector in this subspace, which is known because
all pairs of orthogonal vectors are known. Suppose such a vector is
$c|\psi_1 \rangle+d|\psi_i \rangle$. Then the orthogonality implies
$$c\langle \psi_1|\psi_1\rangle+b^*d\langle \psi_i|\psi_i\rangle = 0 .
\eqno(3.2)$$
Since both $c$ and $d$ cannot be zero, $d\ne0$. Therefore, (3.2)
determines $\langle \psi_i|\psi_i \rangle$ in terms of $\langle
\psi_1|\psi_1\rangle$ for $i=2,3,...N+1$. Therefore if we assign a
value to $\langle \psi_1|\psi_1\rangle$ then this determines the
inner product. This assigned value can be changed by a real positive
factor. There is also an ambiguity of complex conjugation of the
inner product due to two different conventions of defining the inner
product $\langle .|.\rangle$ as being linear in the second slot and
antilinear in the first slot or vice versa. But this ambiguity is
eliminated by fixing the convention to be the former one. Hence the
inner product is unique up to a real positive
factor.

Also from the variation of $\alpha$ on $\cal P$ we can determine the
critical points, which are eigensubspaces of $A$. The values of
$\alpha$ at these points are the corresponding eigenvalues. (To
determine the critical points of an observable with distinct
eigenvalues only the topology of ${\cal P}$ reconstructed earlier is
needed; the differential and projective structures, and the  metric on
$\cal P$ are not needed.) From its eigenvaues and eigenvectors, of
course, the Hermitian operator $A$ in a given $\cal H$ is uniquely
determined.

Once the inner product is determined from the observed values in a
protective measurement via the above theorem, the Fubini-Study
metric [5] on ${\cal P}$ is also determined. This metric is
independent
of the ambiguity of the real positive factor in the construction of the
inner product in ${\cal H}$ mentioned above. It is clear from the
proof of the above theorem that in order to determine all pairs of
orthogonal vectors and thereby reconstruct the Hilbert
space structure, we need to know only the expectation values of
yes-no observables, i. e. Hermitian operators with two distinct
eigenvalues. Indeed, it is sufficient to know the expectation values of
rank one Hermitian operators with eigenvalues $0$ and $1$ for
this reconstruction.

The crucial role played by the protective experiment is in
determining the value of $\alpha$ for each pair of state and
observable, deterministically using a single system. This is in sharp
contrast with the usual measurement in which the state collapses
into one of the eigenstates of the observable, and therefore $\alpha$
is meaningful for a single system only only when it is in an
eigenstate. At other points in ${\cal P}$ the usual interpretation of
${\alpha}$ is that it is "the expectation value", i. e. a {\it statistical}
average which can be given physical meaning only by having an
ensemble of identical systems even for a fixed state and observable.

It would be strange for a quantity to have ontological meaning at
certain points on ${\cal P}$ but epistemological meaning in all other
points on ${\cal P}$. On the other hand, protective measurement
gives
ontological meaning to $\alpha$ at {\it all} points in ${\cal P}$, and
therefore satisfies the fundamental requirement of continuity in
physics. The above theorem which enables the linear structure, the
inner product, observables in ${\cal H}$ and
the Fubini-Study metric on
${\cal P}$ to be reconstructed from the observed values of $\alpha$
therefore gives the same ontological meaning to these structures as
well. The usual measurement which gives meaning to a state by the
eigenvalues of observables for which the state is an eigenstate needs
to assume the linear structure of ${\cal H}$. The additional meaning
that it gives to the state through the probability amplitude for
transition presupposes the inner
product. But protective measurement need not presuppose any of
these structures, because of the above theorem.

As an example, consider the special case of the ideal Stern-Gerlach
protective experiment for a spin 1/2 particle described in Sec.~2.
The
spin state is protected by a large magnetic field in the direction ${\bf
m}$ of the spin. Here ${\cal H}$ is two dimensional and ${\cal
P} = CP(1)$
is a real sphere whose points may be identified with the possible
directions of the spin. It is not required here that the angles between
the different directions are measurable but only that the different
directions are distinguishable.

Suppose that the ``magnetic field'' is in the
direction of ${\bf n}$. The time averaged average force on the
particle may be
protectively measured by the change in momentum of its center of
mass (apparatus). Here, ${\bf m}$ is known, but ${\bf n}$ may be
unknown before the experiment, which is the reverse of the situation
considered before [2]. But ${\bf n}$ may be experimentally
determined
by the property that this force is maximum when ${\bf m}=\bf n$.

Denote the magnitude of the observed change in momentum for a
given pair of state
and observable, defined above, by $\alpha_{\bf n}({\bf m})$. The
two values of ${\bf m}$ for which the $\alpha$ is maximum and
minimum correspond to orthogonal rays. Now vary ${\bf n}$ and
thereby determine all  pairs of orthogonal rays. From this the inner
product is reconstructed uniquely up to a constant multiple. The
Fubini-Study metric obtained from the inner product is the usual
metric on the sphere $\cal P$. And each pair of orthogonal rays are
opposite points on this sphere.

For a given ${\bf n}$ the values of ${\bf m}$ for which
$\alpha_{\bf n}({\bf m})$ takes its the maximum and minimum
values correspond to the eigenvectors of the observable being
measured. The corresponding values of $\alpha_{\bf n}({\bf m})$,
namely $\mu$ and $-\mu$, are its eigenvalues. This determines the
observable to be the Hermitian operator
$\mu {\bf \sigma}\cdot {\bf n}$
so that $\alpha_{\bf n}({\bf m})=\mu \langle \psi|{\bf \sigma}\cdot
{\bf n}|\psi \rangle$ where $|\psi \rangle$ is the normalized state
satisfying (2.8).

To recapitulate, by looking at the observed expectation values
"locally" at a given point on ${\cal P}$, this point can be determined
as
a ray in ${\cal H}$. By looking at the expectation values "globally"
over
all of ${\cal P}$, the linear structure and the inner product in
${\cal H}$, the
Fubini-Study metric in ${\cal P}$, and the observables can be
determined.
If the system is a charged particle, then its wave functions undergoes
local gauge transformations. These transformations are
diffeomorphisms on ${\cal P}$ whose points  are {\it not} invariant
under them. However, the local gauge transformations are obviously
linear transformations and therefore they preserve the projective
geometry in ${\cal P}$. In addition they are unitary transformations
in
${\cal H}$ because they preserve the inner product. Therefore the
inner product in ${\cal H}$ and the Fubini-Study metric in ${\cal P}$
are
invariant under these transformations.

This is analogous to general relativity in which, because Einstein's
field equations are generally covariant, the metric is determined only
up to diffeomorphisms on space-time. So, the points of space-time, to
be regarded in any invariant manner, should be determined by
intersecting trajectories that are specified by metrical relations
which alone are invariant. The same remark
applies to the points in ${\cal P}$. So, in the same way that for a
given
gravitational field, it is first necessary to obtain the metric on space-
time before we can physically specify the points, it is necessary to
construct the Fubini-Study metric in ${\cal P}$ before the physical
states can be determined. Hence, to give ontological meaning to the
physical states by means of protective measurements on a single
system, it is necessary to be able to construct the Fubini-Study
metric on ${\cal P}$, which has been shown to be possible in this
section.

These considerations suggest that the Fubini-Study metric may be
dynamical like the space-time metric in the sense of not being fixed
to have constant curvature for all physical situations. But it should be
kept in mind that the
space-time metric is intimately associated with the equivalence
principle in general relativity which may be stated as follows: If the
same gravitational field is probed by different test particles, the
same space-time geometry is obtained. This principle is {\it
complemented} in quantum theory by the following principle which
is implicit in the above theorem: If the same particle undergoes
different interactions leading to measurement on it of different
observables, the same geometry on the set of physical states of the
particle is obtained.
\vtwo\noindent
{\bf {4. PROTECTIVE MEASUREMENT OF THE

\parindent=1.25em

SPATIAL  WAVE FUNCTION, USING THE

GRAVITATIONAL FIELD}}
\vone

All particles interact gravitationally. Therefore a universal method
for determining the wave function of any particle in principle is by
doing a protective measurment  using its gravitational field. Since the
wave function does not collapse during the protective measurement,
 its gravitational field may be obtained, as will be shown below,
from the semi-classical Einstein's field equations
$$G^{\mu \nu}= 8 \pi G\langle \psi  |{\hat T}^{\mu \nu}|\psi \rangle
\eqno(4.1)$$
where ${\hat T}^{\mu \nu}$ is the energy-momentum
tensor operator, $G^{\mu \nu}$ is the Einstein's tensor formed from
the metric $g_{\mu \nu}$,
and G is Newton's gravitational constant. In principle, by protectively
measuring $g_{\mu \nu}$ the quantum state
$|\psi \rangle$ can be determined via (4.1).
 But (4.1) is severely violated in the usual measurements during
which the wave function collapses [8].

I shall prove in a simple case that the semi-classical gravitational
field equations are
a good approximation for the above mentioned protective
measurement whereas they are not for the usual measurement.
The generalization of this result for an arbitrary wave function will
then be straightforward. Consider two identical spherical containers
separated by a distance $2a$ that is large compared to their radius.
Suppose that the normalized wave function of a particle
with mass $M$ at time $t=0$ is
$$\psi({\bf x},0)=\psi_1({\bf x},0)+
\psi_2({\bf x},0) , \eqno(4.2)$$
where $\psi_1({\bf x},0)$ and $\psi_2({\bf x},0)$
represent the ground states
inside the two containers. The norms of $\psi_1$ and $\psi_2$
need not be equal. At this instant, a test particle
with mass $m$ is observed to be approximately at rest at the
midpoint
of the line joining the centers of the two containers. Choose a
Cartesian coordinate system whose origin is at this midpoint and its
z-axis along the line joining the centers of the two containers. Then
the combined wave function
is $\psi({\bf x},0)\phi ({\bf r},0)$,
where $\phi ({\bf r},0)$ is a normalized wave
function that is localized
in a neighborhood small compared to $2a$ with the expectation
values of its momentum and position
coordinates zero, and the uncertainty in the momentum small.

This test particle has very low energy and interacts with
the weak gravitational field of $M$. Therefore, Newtonian gravity is
a good approximation in this case. The Hamiltonian for the combined
system is
$$H=-{\hbar^2\over 2M}\nabla_{\bf x}^2-{\hbar^2\over 2m}
\nabla_{\bf r}^2+U({\bf x})-{GMm\over |{\bf r}-{\bf x}|}\eqno(4.3)$$
where $U$ represents the potential that confines $M$ into the two
containers.

The subsequent motion of the combined wave function of the two
particles may be written
$$\chi({\bf r},{\bf x},t)=\psi_1({\bf x},t)\phi_1({\bf r},t)+
\psi_2({\bf x},t)\phi_2({\bf r},t) .\eqno(4.4)$$
The action of the interaction potential, i.e. the
last term of (4.3), on the wave function may be approximated by
taking into account the localization of $\psi_1$, $\psi_2$,
$\phi_1$, and $\phi_2$, and
performing a Taylor expansion to first order in the relative
coordinates $({\bf r}-{\bf x})$:
$$\eqalignno{
-{GMm\over |{\bf r}-{\bf x}|}\psi_i({\bf x},t)\phi_i({\bf r},t)=-
{GMm\over a^2}(2a+(-1)^iz)&\psi_i({\bf x},t)\phi_i ({\bf r},t) ,\cr
                          &\quad i=1,2 &(4.5)\cr
}$$
where $z=r_3-x_3$. Now rewrite the combined wave function
(4.4) in the form
$$\chi=\pmatrix{\psi_1\phi_1\cr\psi_2\phi_2\cr}.
\eqno(4.4')$$
Then, on subtracting the constant term $-2GMm\over a$ from the
Hamiltonian, (4.5) becomes
$${GMm\over a^2}z\sigma_z\chi ({\bf r},{\bf x},t)\eqno(4.5')$$
where $\sigma_z$ is the diagonal Pauli spin matrix.

Now the transition amplitude for excitation of the ground state in
each box due to the weak gravitational interaction is negligible.
Therefore, we can write $\psi_i({\bf x},t)=\beta_i(t) \psi_{ig}({\bf
x})$, where $\psi_{ig}, i=1,2$, are the normalized ground state wave
functions in the two boxes. It follows that, in the present
approximation
$M$ may be treated as a two
state system and the wave function of the two particles represented by a two
component spinor function of ${\bf r}$ and $t$.
Then $(4.5')$ is like the
interaction term when a spin-1/2 particle interacts with an ideal
Stern-Gerlach magnetic field whose direction and
inhomogeneity are in
the z-direction. Here, $\psi_{ig}$ is like a spin state and $\phi_i$ is like
a center of mass wave function of the spin-1/2 particle.

Each
of the two states in the entanglement (4.4)
corresponds to $M$ being in
one of the two containers with $m$ moving towards it. Hence, as in
the Stern-Gerlach experiment, if we observe which trajectory $m$
actually takes by letting it interact with a macroscopic system, a
collapse of the wave function takes place so that $m$ would be
observed to move towards one of the two containers as if the particle
$M$ is entirely in it, and not a
superposition of the two possibilities as predicted by Schr\"odinger's
equation. This is also like how in the double slit experiment if we
observe through which slit the particle went then this
destroys the interference pattern. Therefore, for the usual
measurement, the
semi-classical field
equation (4.1) is not valid,
because (4.1) predicts that the gravitational field experienced by
$m$ is the average of the gravitational fields due to $\psi_1$ and
$\psi_2$ in the two containers. Hence, the gravitational field must be
quantized.

Now let us protect $\psi$ by connecting the two containers by a tube
so that there is tunneling between the two boxes. This modifies (4.3)
by adding to it a Hermitian operator $H_P$ that represents the effect
of tunneling. Since a constant term may be subtracted from a
non relativistic Hamiltonian, $H_P$ may be assumed to be a trace
free operator in the two dimensional reduced Hilbert space of $M$
introduced above. I.e. $H_P=B_o{\bf \sigma}\cdot {\bf m}$, where
${\bf m}$ is a unit vector. $B_0$ is analogous to the protecting
magnetic field in the Stern-Gerlach experiment studied in Sec.~2.
The state $\psi$ being protected is a non degenerate eigenstate of
$H_P$. The protective measurement is made adiabatically so that
(2.3) is satisfied. Then $\psi$ remains as an eigenstate of $H_P$ so
that the combined wave function continues to have
 the product form $\psi\phi$.
No entanglement takes place, unlike (4.4). But it follows
from (2.6) and $(4.5')$that $m$ experiences a force in the z-direction
$$f=-{GMm\over a^2}\langle\chi|\sigma_z|\chi\rangle .\eqno(4.6)$$
This force is as if the gravitational field experienced by $m$
due to $M$ is the average of the gravitational fields due to $\psi_1$
and $\psi_2$ in the two containers, i.e.
the semi-classical equations (4.1) are valid to a good approximation
for this protective measurement. This is very different from the
usual measurement discussed above.

More generally, consider an arbitrary wave function $\psi$ that is
protectively measured using its gravitational field, under the same
approximations as above. The wave function of the combined system is then in
the product form $\chi({\bf r},{\bf x},t)=\psi ({\bf x},t)\phi ({\bf r},t)$,
where it is assumed that $\phi$ is localized around ${\bf r}={\bf r}_0$.
Therefore the force experienced by $m$ is
$${d\over dt}\langle \chi |-i\hbar \nabla_{\bf r}|\chi\rangle =GMm\int d^3x
\psi^*({\bf x},t){{\bf r}_0-{\bf x}\over |{\bf r}_0-{\bf x}|^3}\psi ({\bf x},t)
.$$
I.e. the force is like as if the mass $M$ is spread out over the entire wave
function $\psi$. This is unlike the usual measurement in which the measurement
on $\phi$ would collapse $\psi$ so that the force is like as if $M$ is
localized around one point, which was illustrated above in the special case of
$M$ in the two containers. Hence, the semi-classical gravitational equations
can be assumed to be applicable for the field of a general wave function that
is protected.

The gravitational
field, of course, would be weak in such an
experiment. Therefore  a coordinate system can be chosen such that
the classical gravitational field is
$g_{\mu \nu} = \eta_{\mu \nu} + h_{\mu \nu}$, where $\eta_{\mu
\nu}$ is the Minkowski metric and $h_{\mu \nu} << 1$. Define
$h_\mu  = ({h_{00}\over 2}, h_{01} , h_{02} ,h_{03})$. Then ${\bf g} =
-\nabla h_0$ is the acceleration due to gravity. Suppose
 ${\bf h}= - (h_{o1} , h_{o2} ,h_{o3})$. Then ${\bf H} = \nabla \times
{\bf h}$ is twice the
angular velocity of the coordinate system
relative to the local intertial frame, in units in which $c=1$.
Assuming all energies to be low,
the classical gravitational field equations
may be approximated by [9]
$$\nabla \cdot {\bf g}=-4\pi Gj^0,  \nabla \times {\bf g}={\bf 0},
\nabla \cdot {\bf H}=0, \nabla \times {\bf H}=4(-4\pi G
{\bf j}+{\partial
{\bf g}\over \partial t})\eqno(4.7)$$
where $j^\mu = (j^0 ,{\bf j})$ is the mass 4-current density. (4.7) is
more general than what is implied by (4.3) because of the inclusion
of the gravitomagnetic field $\bf H$ due to the motion of the source.

In view of the above arguments, the sources of the gravitational field
in (4.7)
due to a single particle of mass $M$ that is being a protectively
measured are to a good
approximation the quantum
mechanical mass intensity $j^0$ and mass
current density $\bf j$, defined by
$$\eqalign{j^0(\bf x) &= M\psi^*({\bf x})\psi({\bf x}),\cr
{\bf j}({\bf x}) &= {i\hbar M\over 2}\bigl[\{\bigl(\nabla +i{M\over
\hbar}{\bf h}\bigr)\psi^*\}\psi({\bf x})
-\psi^*\bigl(\nabla
-i{M\over \hbar}{\bf h}\bigr)\psi({\bf x})\bigr],\cr
&= j^0({\bf x})\{\hbar \nabla \theta ({\bf x})-M{\bf h}({\bf
x})\}.\cr}\eqno(4.8) $$
This makes (4.7) an approximation of (4.1) with $j^\mu$ replacing
the components $\langle \psi |{\hat T}^{\mu 0}|\psi \rangle$
of the energy-
momentum tensor. The minimal coupling of
${\bf h}({\bf x})$ in (4.8) is physically due to $\bf j$ being the
current density relative to the present coordinate system. The
geodesic equation for a freely falling test particle in the gravitational
field is approximated by
$${d{\bf v}\over dt}={\bf g}+{\bf v}\times {\bf H}\eqno(4.9)$$

Consider a quantum mechanical particle which is in a non degenerate
eigenstate $\psi ({\bf x})$ of the Hamiltonian. A protective
measurement of its wave function may be made as follows: Shoot
test particles to move in the gravitational field of $\psi ({\bf x})$
such that a) the particles travel slowly so that they cover a distance
in which the gravitational field varies appreciably in time
$T>>{\hbar\over \Delta E}$, where $\Delta E$ is the smallest of the
magnitudes of the energy differences between $\psi$ and the other
energy eigenstates, and b) the potential energy of interaction
between each test particle and the given particle is at all times small
compared to $\Delta E$. Then the interaction is adiabatic and $\psi$
does not become entangled with the probe of the
gravitational field.

By observing the acceleration of the test particles,
using (4.9), ${\bf g}$ and ${\bf H}$ fields due to $\psi$ can be
determined. Then by solving
(4.7), $j^{\mu}$ can be determined. From $j^{\mu}$, $\psi({\bf x})$
can
be reconstructed like in the electromagnetic case [1,2]. The
observation of the test particle trajectories, which ultimately
requires interaction with a macroscopic apparatus, does not collapse
$\psi$ because of the absence of entanglement. Therefore, $\psi$ can
be determined in this manner, deterministically, for a single particle
up to gravitational gauge transformations.

As a specific example, consider a particle of mass $M$
inside a circular narrow tube of radius $R$. This is analogous to the
experiment described in fig.
 3 of ref. [2], which has the drawback that it would work only for a
charged particle unlike the present experiment. Then the state with a
given angular
momentum is degenerate with the state with angular momentum in
the opposite direction. We can remove this degeneracy by
transforming to a frame rotating with angular velocity
${\bf \Omega}$
about the axis of the tube. Then, in this rotating frame, the
Hamiltonian is
$$H_0 = {1\over 2M} ({\bf p}-M{\bf h})^2 +V({\bf x}) \eqno(4.10)$$
where ${\bf h} = {\bf \Omega}\times {\bf x}$ and $V({\bf x})$
represents the centrifugal energy and
 the interaction of the particle with the
walls of the tube. The low lying eigenvalues of (4.10) are to a good
approximation
$$E_n = {(n\hbar-MR^2 \Omega)^2\over 2MR^2}+ K  ,\eqno(4.11)$$
where n is an integer, which corresponds
to the angular momentum (classically
$Rp$) having quantized values $n\hbar$, and $K$
 is a constant. Inspection of (4.11) shows
that these energy eigenvalues are all distinct provided
$$MR^2 \Omega \ne s{\hbar\over 2} ,\eqno(4.12)$$
for any integer $s$, which is assumed to be the case.

Choose cylindrical coordinates about the axis of the tube: $r, z ,\phi$
in the inertial frame and $r,z,\phi ^\prime$ in the rotating frame.
Then,
$$ \phi ^\prime = \phi -\Omega t .\eqno(4.13)$$
Make now a protective measurement of the projection operator
$P_{r, z ,\phi ^\prime}$ $:= |r, z ,\phi ^\prime\rangle \langle r, z
,\phi ^\prime |$. This gives the value
$$\langle \psi |P_{r, z ,\phi ^\prime}|\psi \rangle = \psi^*(r, z ,\phi
^\prime)\psi (r, z ,\phi ^\prime)=j^0,\eqno(4.14)$$
which is the intensity of the wave $\psi$. Then, protectively measure
${\bf A}={1\over 2}(P_{r, z ,\phi ^\prime}{\bf \Pi} +
{\bf \Pi} P_{r, z ,\phi ^\prime})$,
where $\bf \Pi$ is the kinetic momentum operator. This gives the
current density $\bf j$ [1,2].
These two measurements may be made as follows: shoot particles
near the tube and from their trajectories determine $\bf g$ and
$\bf H$ using (4.9). By solving (4.7), $j^0$ and $\bf j$ can then be
determined. From $j^0$ and $\bf j$, the wave
function $\psi$ can be reconstructed up to gauge
transformations using (4.8).

\vtwo\noindent
{\bf{5. PROTECTIVE MEASUREMENT OF
\parindent=1.25em

DEGENERATE STATES}}
\vone

The protective measurement of the wave function inside the tube,  in
Sec.~4, raises the following paradox: In the rotating frame, the
wave
functions are non degenerate, which enabled the protective
measurement to be possible. But in the inertial frame each state is
degenerate with the state with the opposite angular momentum
which would make it appear that protective measurement is not
possible. The resolution of this paradox below will show a way of
making protective measurements of degenerate states by using time
dependent Schr\"odinger observables.

First of all if we wish to determine the intensity of the wave function
in the inertial frame for a single particle by means of time
independent observables then we need to measure protectively the
expectation value of
$$P(r, z,\phi)=|r,z,\phi\rangle \langle r,z,\phi | ,$$
which is the projection operator of an arbitrary point that is at rest
in the  inertial frame. Since this is periodic in  $\phi$ with period
$2\pi$,
$$P(r, z,\phi)= \sum_{m=-\infty}^{+\infty} P_m(r,
z)e^{im\phi},\eqno(5.1)$$
where the operators $P_n$ satisfy $P_{-m}=P_m^{\dagger}$ for every
m.
Let $|n\rangle$ be the eigenfunction of  the Hamiltonian in the
inertial frame with angular momentum $n\hbar$, where n is a  non
zero integer. Then $|n\rangle$ and $|-n\rangle$ are degenerate. Also,
one can write
$$\langle r,z,\phi|n\rangle =\psi _n(r,z)e^{in\phi}.$$
It follows that
$$\langle -n|P(r, z,\phi)|n\rangle =\psi _n^*(r,z)\psi
_n(r,z)e^{i2n\phi}\ne 0.\eqno(5.2)$$
Therefore if we attempt to  measure the intensity of $|n\rangle$ by
an experiment in which $A=P(r, z,\phi)$ in (2.1), then $A$ would
cause transition from $|n\rangle$ to $|-|n\rangle$. Hence a protective
measurement of $P(r, z,\phi)$ is not possible basically because this
operator connects the two degenerate states $|n\rangle$ and $|-
n\rangle$.

Now, let us look at the same experiment in the rotating frame. The
states $|n\rangle$ and $|-n\rangle$ transform to non degenerate
states $|n^\prime\rangle$ and $|-n^\prime\rangle$ whose energy
difference is given by (4.11) to be
$$E_{-n} -E_n =2n\hbar \Omega .\eqno(5.3)$$
But the term in the sum (5.1) which contributes to (5.2) is
proportional  to $e^{i2n\phi}$. In the rotating frame this term
transforms to $P_n(r, z)e^{i2n(\phi^\prime+\Omega t)}$ in view of
(4.13). The latter operator has the right time dependence to cause
transition from $|n^\prime\rangle$ and $|-n^\prime\rangle$ in time
dependent perturbation theory because of (5.3). Therefore the
measurement of $P_{r, z ,\phi}$ cannot be done adiabatically in the
rotating frame, which is consistent with the fact that it is not a
protective measurement in the inertial frame.

On the other hand, it was shown in Sec.~4 that
$P_{r, z ,\phi^\prime}$ can be protectively measured
 in the rotating frame. In
the inertial frame, this observable is $P_{r, z ,\phi -\Omega t}$
and is
periodic in the variable $\phi -\Omega t$ with periodicity $2\pi$.
Therefore,
$$P(r, z,\phi^\prime)= \sum_{m=-\infty}^{+\infty} P_m(r,
z)e^{im(\phi -\Omega t)},\eqno(5.4)$$
Hence, by time dependent perturbation theory, none  of the terms in
(5.4) with non zero frequency could cause a transition between the
degenerate states $|n\rangle$ and $|-n\rangle$. The zero  frequency
term also  does not cause such a transition because the identity
analogous  to (5.2) here  implies that the only term in the sum in
(5.4) which makes a non zero contribution to the matrix element is
the $m=2n$ term. Therefore,
$$\langle-n|P_0(r, z)|n\rangle =0.$$

To summarize, it has been shown both in the inertial frame and the
rotating frame that
$P_{r, z ,\phi}$ cannot be protectively measured whereas $P_{r, z
,\phi ^\prime}$ can be protectivley measured. In this example, the
current density (4.8) relative to the rotating frame is also
protectively measured. But this is different from the current density
relative to the inertial frame which is obtained from (4.8) by setting
${\bf h} = \bf 0$. Again the latter current density cannot be
protectively measured, which does not contradict with the protective
measurement of the former because they are different observables.

The above example in which, in the inertial frame, a degenerate state
was protectively measured using a time dependent observable
suggests a generalization of protective measurement which enables
measurement of degenerate states. Suppose each eigenstate of a
complete set of mutually orthogonal eigenstates of the Hamiltonian
$H_o$ of a system is denoted by $|k\rangle$ and the corresponding
eigenvalue by $E_k$. Suppose, $A(t)$ is a time dependent of
observable. It is sufficient to consider an arbitrary Fourier
component of $A(t)$ given by $A_\omega(t) =
A_\omega(0)sin\omega t$. Then the state that was $|m\rangle$ at
$t=-\infty$ will have component along $|k\rangle$ at time $t$ which
is given to first order by [3]

$$a_k(t)=-{\langle k|A_\omega(0)|m\rangle\over i\hbar}
\left\{ {e^{i(\omega_{km}+\omega)t}-1\over \omega
+\omega_{km}}+{e^{i(\omega_{km}-\omega)t}-1\over \omega
-\omega_{km}}\right\} ,\eqno(5.5)$$
where the Bohr frequency $\omega_{km} = {E_k-E_m\over\hbar}$.

The amplitude (5.5) is significant when $\omega_{km}=|\omega|$.
But it is negligible when
$$\left|\omega -|\omega_{km}|\right| \gg {|\langle
k|A_\omega(0)|m\rangle|\over
\hbar } , \eqno(5.6)$$
for every $k$. Then to a good approximation, the system remains in
the original
state $|k\rangle$. No entanglement or collapse. Suppose (5.6) is valid
for each
Fourier component of $A(t)$. Then a protective measurement of
$A(t)$ can be performed to obtain a time average of
$\langle m|A(t)|m\rangle$. In particular this may be done when
$|m\rangle$
is degenerate. Also, (5.6) is valid in the special case when
$\langle k|A_\omega(0)|m\rangle =0$. By measuring a sufficient
number of
time dependent observables $|m\rangle$ can be reconstructed up to
gauge transformations.

To conclude, we have the following result: A protective
measurement of a time dependent observable $A(t)$ can be made on
a given eigenstate $|m\rangle$ provided each term in the Fourier
transform of $A(t)$ in time has frequency $\omega$ that is
significantly different from $|\omega_{km}|$ in the sense of (5.6) for
every $k\ne m$.

The previously studied protective measurements of non degenerate
eigenstates of the Hamiltonian [1,2] are special cases of this result:
The time dependence of the observable in the latter cases was
through $g(t)$ that was required to be slowly varying so that its
Fourier transform did not contain frequencies which would cause
transition from the given eigenstate to eigenstates with different
energies. Also, in order for the zero frequency  component of this
observable not to cause a transition, it was required that the given
eigenstate was non degenerate. But neither of these conditions,
namely adiabaticity and non degeneracy, is required in the more
general protective measurement described by the above result
and illustrated by the above example.
\vtwo\noindent
{\bf{6. EXTENSION TO MANY-PARTICLE SYSTEMS}}
\vone

The concept of protective measurement will now be extended to a
two particle system in non relativistic quantum mechanics. If the
system is in a product of two states $|\psi_1\rangle |\psi_2\rangle$
then this is easily done by protectively measuring each state of the
individual systems. But this is not possible when the system is in an
entangled state because neither particle is then in a unique state that
can be protected. If a protective measurement is attempted on one of
the particles, then this would put it in a mixed state described by a
density matrix that is diagonal in the eigenstates of the protecting
Hamiltonian. And the different
possible outcomes can only be predicted probabilistically.

First consider the example of a system of two spin 1/2 particles
which can be in a singlet $S = 0$ state or a triplet $S=1$ state, where
${\bf S} = {\bf S}_1 + {\bf S}_2={1\over 2}({\bf \sigma}^{(1)} +
{\bf \sigma}^{(2)})$  is the total spin. If we try to protect it just by a
large magnetic field ${\bf B}_0$ say in the z-direction, as was done
for the one particle case, then the $|S = 0\rangle$ state is degenerate
with the $|S = 1, S_z = 0\rangle$ state. So, introduce an additional
interaction proportional to $S^2$. This may be obtained physically
 by a spin-spin interaction energy proportional to
$${\bf S}_1\cdot {\bf S}_2  =
{1\over 2} S^2 - {3\over 4} \eqno(6.1)$$
and subtracting away the constant term from the Hamiltonian.
Assume therefore the Hamiltonian
$$H_0 = {{\bf p}_1^2\over 2M}  + {{\bf p}_2^2\over 2M} - \mu B_0
{\bf S}\cdot {\bf m}  - \lambda {\bf S}^2 , \eqno(6.2)$$
where ${\bf p}_1$ and ${\bf p}_2$ are the momenta of the particles,
${\bf m}$ is a fixed unit vector, $B_0$ and $\lambda$ are assumed
to be large.

$H_0$ is non degenerate, and therefore any of its eigenstates may be
protectively measured. The eigenstates are the singlet $S=0$ state
and the three triplet $S=1$ states. All but one of the six pairs of
eigenstates may be
distinguished using the general scheme for making local
measurements of non local observables due to Aharonov
and Albert [10].
The basic idea in this scheme is to prepare the apparatus in a
suitable quantum state before it interacts with the system and infer
a non local property of the system by making local measurements
on the apparatus after the interaction. I shall choose as the apparatus
the center of mass coordinates ${\bf x}^{(1)}$ and ${\bf x}^{(2)}$ of
the two particles. Prepare the apparatus so that, at time $t=0$, it is in
the state for which
$$p^{(1)}_i + p^{(2)}_i =0,    x^{(1)}_i -x^{(2)}_i =0 . \eqno(6.3)$$
i. e. the wave function of the two particles is an eigenstate of these
six observables with eigenvalue zero. This is possible because these
observables commute with one another. This is the state of two
particles which are Einstein-Podolsky-Rosen [11] correlated in
position and momentum coordinates.

Now we let this system interact during the time interval [0,T] with a
Stern-Gerlach apparatus with a magnetic field $B_i = L_{ij}x_j$, using
the summation convention, where $L_{ij}$ is a non singular,
symmetric traceless matrix. Then $div {\bf B} = 0, curl {\bf B} = {\bf
0}$. The total Hamiltonian $H = H_0 + H_I$,
where the interaction Hamiltonian
$$H_I = - \mu g(t) L_{ij}(x^{(1)}_j \sigma^{(1)}_i +
x^{(2)}_j\sigma^{(2)}_i) . \eqno(6.4)$$
Then, denoting the state of the combined system by $|\psi\rangle$,
$${d\over dt} \langle \psi|p^{(1)}_i|\psi\rangle = L_{ji}\mu g \langle
\psi|\sigma^{(1)}_j|\psi\rangle, {d\over dt} \langle
\psi|p^{(2)}_i|\psi\rangle = L_{ji}\mu g \langle \psi|\sigma
^{(2)}_j|\psi\rangle \eqno(6.5)$$
Therefore, using (6.3),
$$\langle \psi|p^{(1)}_i + p^{(2)}_i|\psi\rangle_{t>T} = \mu L_{ji}
\langle \psi|\sigma^{(1)}_j +\sigma^{(2)}_j|\psi\rangle , \eqno(6.6)$$
where the normalization (2.7)
is assumed.

Now determine $\langle \psi|p^{(1)}_i|\psi\rangle_{t>T}$ and $\langle
\psi|p^{(2)}_i|\psi\rangle_{t>T}$ by local measurements and thereby
determine (6.5). Since $L_{ij}$ is non singular, this determines
$\alpha _j =\langle \psi|\sigma^{(1)}_j +\sigma^{(2)}_j|\psi\rangle$. It
may be noted that it is not possible to determine $\langle
\psi|\sigma^{(1)}_j|\psi\rangle$ for example by these measurements
via (6.5) because we do not know
$\langle \psi|p^{(1)}_i|\psi\rangle_{t=0}$.
If $|\psi\rangle = |S=1, {\bf S}\cdot {\bf m}= 1\rangle$ or
$|S=1, {\bf S}\cdot {\bf m} = -1\rangle$ then $\alpha _j = 2m_j$, or
$-2m_j$ respectively.  Hence, in either case the state can be
determined. But if $|\psi\rangle = |S=0\rangle$  or $|S=1, {\bf
S}\cdot {\bf m} = 0\rangle$ then $\alpha_j = 0$. Therefore, these two
states cannot be distinguished by means of the above measurements.
But they
may be distinguished by the measurement of
${\bf \sigma}^{(1)}\cdot {\bf \sigma}^{(2)}$.
But this does not determine {\bf m} and therefore
cannot determine the latter state.

But a measurement that distinguishes the last pair of states, and
determines the last mentioned state, may be performed by a
quadrupole interaction that determines the expectation value of the
observable $A=({\bf S}\cdot {\bf n})^2$, where ${\bf n}$ is a unit
vector in the direction of an external field or in the direction of the
line joining the center of masses of the two particles. It is easy to
show that
$$\langle S=0|A|S=0\rangle =0,\langle S=1, {\bf S}\cdot {\bf m} =
0|A|S=1, {\bf S}\cdot {\bf m} = 0\rangle = sin^2\theta \eqno(6.7)$$
where $\theta$ is the angle between the vectors ${\bf n}$ and ${\bf
m}$. Hence, by varying the direction of ${\bf n}$, ${\bf m}$ and
therefore the state represented by the state vector $|S=1, {\bf S}
\cdot {\bf m} = 0\rangle$  may be determined.

This quadrupole interaction, like (6.1),
is non local for separated particles. However, this
measurement may be performed without violating Einstein causality
by having the two particles sufficiently close to each other so that
they have this interaction. Then when the particles are separated
they would still be in the same spin state which has been
protectively measured.

Not every spin state of course is an eigenstate of (6.2). But since
$SU(3)$ is generated by the three components of the spin vector
${\bf S}$ and the five components of the symmetric traceless
quadrupole tensor
$$Q_{ij} = {1\over 2}(S_iS_j +S_jS_i -{4\over 3}\delta_{ij} I)
\eqno(6.8)$$
by adding appropriate terms formed from (6.8) to (6.2), any spin
state may be a non degenerate eigenstate of $H_0$. e. g. a neutron
and a proton has a tensor interaction which makes the $
|S=0\rangle$ state unbounded, whereas the $|S=1\rangle$ states are
bounded leading to the formation of the deuteron. Each of these
states can be determined
by protectively measuring the above eight
observables.

Instead of starting with ${\bf m}$ unknown and determining it by
protective measurements,  we may begin, as in the approach in
Section
3, by knowing what ${\bf m}$ is but without knowing the linear
structure or the inner product in ${\cal H}$. Then the above
measurements
distinguish between the four possible eigenstates of (6.2). By varying
the direction of ${\bf m}$, the expectation values of the above
observables may be determined for an infinite number of states. The
expectation values of these observables with respect to the
remaining states may also be protectively measured by
appropriately modifying (6.2), as mentioned above. The theorem in
Section3 implies that from these observed values, the entire Hilbert
space structure, its inner product and the observables as Hermitian
operators can be constructed.

This scheme can be extended to a many particle wave function by
adding appropriate terms to (6.2) so that the state is a non
degenerate eigenstate of the Hamiltonian. Then the state can be
protectively measured.

Consider now the following apparant paradox, first brought to my
attention by Mermin [12]. Suppose that two widely
separated particles $a$ and
$b$ are EPR correlated [11] in the non commuting observables $P$
and $Q$.
This  state may be used to argue that the wave function of a single
particle cannot be determined, contrary to what is being assumed
here, as follows: If a measurement of $P$ or $Q$ is performed on $a$
then $b$ would be in the corresponding eigenstate of $P$ or $Q$,
respectivley. Therefore, if the state of $b$ can be protectively
measured for a single  particle then we can infer whether the
measurement on the faraway particle $a$ was of $P$ or $Q$ as soon
as the latter measurement is made. This would enable an observer
making measurements on $a$ to send a binary signal faster than the
speed of light to an observer making protective measurements on
$b$, violating Einstein causality.

This is unlike the case when the measurements are of the usual type
which require an ensemble of particles to determine the wave
function. In the latter case, if the second observer measures $P$,
different eigenvalues of $P$ would be observed for the different
members of the ensemble from which it cannot be determined if the
first observer measured $P$ or $Q$. Because  the reduced density
matrix for $b$ is proportional  to the identity matrix and therefore
does not distinguish between $P$ and $Q$.

But this paradox is resolved as follows: It is possible to protect the
eigenstates of the non commuting variables $P$ or $Q$, but not both.
Suppose the eigenstates of $P$ of particle $b$ are protected by a non
degenerate Hamiltonian $H$ that has these states as its eigenstates.
Then if the first observer measures $P$ on $a$, then the protective
measurement of the second observer would show $b$ to be in the
corresponding eigenstate of $P$. However, if the first observer
measures $Q$ on $a$, then the same experiment performed by the
second observer on $b$ would collapse the state of $b$ (an eigenstate
of $Q$) into one of the possible eigenstates of $P$ which would then
be observed by the protective measurement. Hence this
measurement cannot determine whether the first observer measured
$P$ or $Q$.

For example, let $a$ and $b$ are neutrons, $P=\sigma _z$ and
$Q=\sigma _x$. A large homogneous magnetic field $B_0$ in  the z-
direction acts on $b$ thereby protecting its eigenstates of $\sigma
_z$. If $a$ is measured to be in an eigenstate of $\sigma _z$ then $b$
would be in the corresponding eigenstate of $\sigma _z$. The latter
state can be protectively measured by sending the neutron through
a Stern-Gerlach  apparatus as shown in Sec.~2. But if $a$ is
measured to be in an eigenstate of $\sigma _x$ then $b$ would be in
the corresponding eigenstate of $\sigma _x$, which is an equal
superposition of the two eigenstates of $\sigma _z$. The
 Stern-Gerlach experiment would collapse it into one of
these two eigenstates which is what would be measured by the
protective measurement. Thus it would not be possible
to determine which observable was measured by the first observer
in either case.
\vtwo\noindent
{\bf{7. IS THE WAVE FUNCTION REAL?}}
\vone

First of all what does 'real' mean? Physicists in general would agree
that for an object or concept to be real it is necessary and sufficient
that an experiment can be performed  in
principle to measure it. But there are two ways of making a
measurement: It can be made on a single system as when we
observe a macroscopic system, for example a particular tree which
has no other tree like it. Or a measurement can be made on an
ensemble of identical systems or experiments, and the data
accumulated is used to infer statistically the properties of the
system. This may be done in practice just to eliminate errors due to
imperfect measuring instruments while a precise measurement is
possible in principle on a single system as in classical physics. Or
even with the most perfect instruments an inherent uncertainty may
force us to make statistical measurements as in quantum physics.

The determination by a measurement on a {\it single} system of a
property possessed by the system prior to the
measurement gives
a stronger version of reality to this property, which will
be called {\it ontological reality}, whereas the weaker version of
reality obtained by measurements on an ensemble of similar systems
will be called
{\it statistical reality} here.

 An interesting argument was advanced by Einstein [13] in favor of
the statistical reality and against the ontological
reality of the wave function. While this argument would look
reasonable if the only type of measurement we have is the usual one
that results in the collapse of the wave function, it will be refuted
here using the protective measurement. Einstein considered a
particle in a box whose wave function is the following solution of
Schr\"odinger's equation:
$$\psi = C exp (-i{E\over \hbar}t ) coskx .\eqno(7.1)$$

Einstein then took the classical limit as $m \to 0$ or $\hbar \to 0$
while keeping the energy $E$ constant. In this limit the position
probablility distribution is uniform. But the classical particles
which we have observed so far always have localized states and
cannot be described by such an extended wave function.
However, this
apparant conflict with observation would not occur if the wave
function, to begin with, described an {\it ensemble} of identical
particles instead of a single particle. Then in the classical limit, the
ensemble of particles described by the wave function is uniformly
distributed, and this limit is smooth. Therefore Einstein concluded
that the wave function always describes an ensemble of identical
systems and that quantum mechanics is incomplete because it cannot
describe a single system.

To this Born responded [14] by arguing that Einstein did not specify
the proper initial condition for a macroscopic system. He believed
that the initial condition should be such that  the wave function is
localized. Then, for a large enough mass the spreading of the wave
function would be negligible. And therefore it remains
localized, consistent with observation. Ironically, even though Born
and Einstein disagreed, they agreed on the assumption that a {\it
single} macroscopic particle must necessarily have a localized wave
function (which I believe to be an incorrect assumption within the
framework of standard quantum theory). Their disagreement is due
to Einstein  giving universal reality to all solutions of Schr\"odinger's
equation in the sense of requiring that each of them must describe at
every time a possible physical situation that could exist in
principle. Whereas Born denied that a macroscopic particle can even
in principle have an extended wave function.

In spite of Einstein's argument [13,14] that the principle of
superposition
and the spreading of the wave function over time according to
Schr\"odinger's equation requires that extended wave functions are
admissible even for a macroscopic particle, Born refused to accept
this. Thus Born seemed to deny even statistical reality for an
extended wave function of a macroscopic particle.

The only way Einstein found to give reality to the wave function is
by giving it statistical reality by supposing that it represents an
ensemble of identical systems. This is consistent with the statistical
outcome when the usual measurement is made to observe the
particle leading to the collapse of the wave function. However,  by
means of a protective measurement the extended wave function
(5.1) can  in principle be measured for a single macroscopic particle.
For example,  we can use the gravitational field due to the particle as
in Sec.~4 to determine the wave function.
This would give ontological
reality to the wave function, in disagreement with Einstein's claim
that the wave function can have only statistical reality.

In practice, such a protective measurement is very difficult or
impossible because of two reasons: (i) It is extremely difficult to
isolate a macroscopic particle from its environment so that it may
be regarded as having an extended wave function. If initally the
state of the state of the universe is a product of the particle state and
the environment state then the interactions in a very small period of
time make the two systems hopelessly entangled:
$$|\psi \rangle |\alpha \rangle \to \sum_i  |\psi _i \rangle |\alpha _i
\rangle . \eqno(7.2)$$
In other words, decoherence [15] takes place very rapidly. Here each
$|\psi _i \rangle$ is a localized state of the particle.  This is due to the
fact that the interactions of the particle, which we believe to be
ultimately due to gravity and gauge fields, are represented by terms
in the Hamiltonian that are functions of space-time. Therefore, they
commute with the position operator and hence are
diagonal in the basis
of position eigenstates. This is ultimately the reason why macroscopic
particles look classical in the sense of being in localized states with  a
fairly well defined position.

(ii) Even if the box is emptied thoroughly so that it is almost a
vacuum, and it is insulated so well that the particle is to a good
approximation interacting only with the box so that we have the
state of the form in the left hand side of (7.2), it is very difficult to
protect the particle state because its energy levels are very closely
spaced. Because the condition (2.3) cannot be satisfied in practice.

But even a macroscopic system can {\it in principle} be observed
protecively to have an extended wave function. Hence the common
assumption that Einstein and Born were making that this would not
be possible is not valid in orthodox quantum theory. However,
quantum theory may be modified to incorporate this assumption.
Such a modification has been proposed by Penrose [16] as a solution
to the measurement problem. He supposes that when a system is in a
superposition of two states whose gravitational field differ by about
one graviton then the state should collapse into one of the
superposed states. This hypothesis, which requires abandoning the
linearity of quantum theory when gravity is quantized, implies that a
macroscopic particle will be in a localized state.

It was shown in section 4 that the collapse of the wave function
implies that gravity must be quantized. This suggests that the
collapse of the wave function is related to quantum gravity. Another
attractive aspect of this proposal is due to the fact that the
gravitational field, as described  by general relativity, is intimately
associated with local description in
physics. Since the collapse of the wave function makes the system
look more localized it seems reasonable to associate this collapse with
the gravitational field.

But the collapse of  the wave function implies that gauge  fields also
must be quantized [2]. Moreover, gauge fields also originate from the
locality of the laws of physics [17]. This is clear
from the fact that gauge
fields are minimally coupled to the matter fields which interact with
it analogous to how the gravitational field is minimally coupled to all
matter fields as implied by the principle of equivalence. Also, it is
reasonable to expect that a quantum theory of gravity should be
such that gravity is unified with gauge fields that describe other
interactions. These considerations suggest that Penrose's proposal for
the collapse of the wave function using the gravitational field should
be extended to gauge fields as well.

The criterion for reality given at the beginning of this section implies
that even with protective measurements the wave function cannot
be given reality. Because the wave function $\psi ({\bf x},t)$ cannot
be distinguished from $\lambda \psi ({\bf x},t)$, where $\lambda$ is
any complex number, by means of any known experiments. This is
unlike the real waves of classical physics such as water waves or
sound waves. So, the protective measurements can be used at most
to give reality to the physical states,
which by the theorem in Sec.~3
become points in the projective Hilbert space ${\cal P}$ or rays in the
Hilbert space ${\cal H}$. What is actually observed, and therefore
must
be regarded as real, are the expectation values of the observables
with respect to a normalized state. So, each observed value is
determined by a pair $(\rho,A)$ where $\rho$ is a point in ${\cal P}$
representing a physical state and $A$ is an observable. Hence, the
protective measurements may be regarded equally as giving reality
to the states (Schr\"odinger picture) or observables (Heisenberg
picture).

Rovelli [18]  and Unruh [4] have argued that the protection already
constitutes a measurement of the protecting Hamiltonian on the
system and therefore the subsequent protective measurement, even
though it is done on a single system, does not establish the reality of
its state. This criticism does not apply to the approach originated in
Sec.~3 of the present paper. Here the
 protection is known and in fact
is part of the measurement that defines the state. Nevertheless the
theorem in this section implies that the states acquire an ontological
meaning in a wholistic manner because the physical meaning of
these states is obtained from the values of the observables and the
relations between the states as determined by the inner product,
both of which are obtained from the expectation values which can be
measured in a non statistical manner by protective measurements.

Also, these authors [4,18] have ignored the important role of
protective measurement pointed out before [2], namely that it shows
the {\it manifestation} of the wave function for a single particle as an
extended object. In the usual measurement, the particle state
collapses to a localized state when the measurement is made which
strongly suggests the usual epistemological meaning to the state. In
the Stern-Gerlach experiment the collapse of the wave function takes
place when the system inetracts with the screen, and again this
collapse is accompanied by a spatial localization of the wave function.
But in the example of a particle in two boxes considered in [2] the
wave function manifests itself as being in both boxes when a
protective measurement is made, whereas the usual measurement
would show the particle as being in one of the boxes. Also, the
protection here is provided by tunneling between the boxes and not
by an interaction with a macroscopic system. Therefore, the
emphasis on the protecting magnetic field of the Stern-Gerlach
experiment being macroscopic in order to argue that this constitutes
a measurement [18] is a red herring.

Even if the objective of the measurement is entirely to obtain
information about the state of the system, protective measurement
accomplishes far more than what these authors [4,18] give it credit
for. Knowing the protecting Hamiltonian does not determine which
eigenstate the system is in, and therefore does not constitute a
measurement of the state. In the two particle system considered in
Sec.~6, it was seen how non trivial it is to distinguish between the
different eigenstates of the Hamiltonian even when the protection is
known. In general, the Hamiltonian may have an infinite
number of eigenstates
and therefore determining which eigenstate the system is actually in
by protective measurements give enormous additional information,
even when the protection is completely known.

In conclusion, whatever one's interpretation of quantum theory may
be, it cannot be denied that protective measurement provides a new
approach to quantum measurement theory. Because unlike the usual
measurement which determines the state from the eigenvaues of
commuting observables, protective measurement treats
democratically {\it all} the expectation values of {\it all} the
observables as having the same validity.
The theorem in Sec.~3 shows
that from these values, which are what are determined by
experiment, the very existence of the Hilbert space and all the
structures in the Hilbert space
 may be inferred. It may
therefore not be unreasonable to expect protective measurement to
be an important milestone in our search for quantum reality.

I thank  Yakir Aharonov for useful discussions. I also
thank Ralph Howard for useful
and informative discussions, and Ralph Howard Young for useful
comments on an earlier version of this paper. This work was
partially
supported by NSF grant no. PHY-8807812.

\vtwo
\noindent{\bf REFERENCES}
\vone
\frenchspacing

\parindent=1.65em

\item{1.} Y. Aharonov and L. Vaidman, {\it Phys. Lett. A} {\bf 178,}
38
(1993).
\item{2.} Y. Aharonov, J. Anandan, and L. Vaidman, {\it Phys. Rev. A}
{\bf
47,} 4616 (1993).
\item{3.} L. I. Schiff, {\it Quantum
Mechanics} (McGraw-Hill, New York, 1968), pp. 289--291.
\item{4.}W. G. Unruh, ``The reality and measurement of the wave
function," unpublished.
\item {5.}S. Kobayashi and K. Nomizu, {\it Foundations of Differential
Geometry}, Vol. II (John Wiley, New York, 1969).
\item{6.}J. Anandan, {\it Found. Phys.} {\bf 21,} 1265 (1991), Sec.~6.
\item{7.} See, for example, D. Pedoe, {\it An Introduction to
Projective
Geometry} (Macmillan, New York, 1963), Chap. X.
\item{8.} D. N. Page and R. Geilker, {\it Phys. Rev. Lett.}
{\bf 47}, 979 (1981);
J. Anandan, ``Interference of geometries in quantum gravity,"
 to be published in the J. of Gen. Relativ. and Gravitation.
\item{9.} R. L. Forward, {\it Proc. IRE} {\bf 49}, 892 (1961);
K. S. Thorne,
in {\it Near Zero: New Frontiers of Physics}
(Freeman, New York, 1988), p. 573.
\item{10.} Y. Aharonov and D. Albert, {\it Phys. Rev. D} {\bf 21},
3316
(1980); {\it Phys. Rev. D} {\bf 24}, 359 (1981).
\item{11.}A. Einstein, B. Podolsky, and R. Rosen, {\it Phys. Rev.} {\bf
47,}
777 (1935).
\item{12.} N. D. Mermin, private communication.
\item{13.} A. Einstein, in {\it Scientific Papers Presented to max
Born}
(Oliver  \&
Boyd, London, 1953).
\item{14.} M. Born, {\it The Born-Einstein Letters} (Walker,
 New York, 1971).
\item{15.} For detailed treatments of decoherence in the context of
the measurement problem, see H. D. Zeh, {\it Phys. Lett. A}
{\bf 172,}189
(1993), and the references therein.
\item{16.} R. Penrose in {\it 300 Years of Gravity}, S. W.
Hawking
and W. Israel, eds. (Cambridge University Press, Cambridge,1987).
R. Penrose, {\it The Emperor's New
Mind} (Oxford University Press, Oxford,  1989), Chap. 8.
\item{17.} See, for example, J. Anandan, {\it Phys. Rev. D}
{\bf 33,} 2280 (1986).
\item{18.} C. Rovelli, Comment on the paper ``Meaning of the wave
function," unpublished.

\bye